\documentclass[preprint, superscriptaddress, showpacs, longbibliography]{revtex4-1}
\usepackage{color, graphicx}
\usepackage{amssymb,amsmath, amsfonts, bm}
\usepackage{bbold}

\begin{document}

\title{Triple point semimetal and topological phase transitions in NaCu$_{3}$Te$_{2}$}

\author{Yunyouyou Xia}
\author{Gang Li}
\email{ligang@shanghaitech.edu.cn}
\affiliation{\mbox{School of Physical Science and Technology, ShanghaiTech University, Shanghai 200031, China}}

\begin{abstract}
Quasiparticle excitations of free electrons in condensed-matter physics, characterized by the dimensionality of the band crossing, can find their elementary-particle analogs in high-energy physics, such as Majorana, Weyl, and Dirac fermions. 
While crystalline symmetry allows more quasiparticle excitations and exotic fermions to emerge. 
Using symmetry analysis and {\it ab-initio} calculations, we propose that the 3D honeycomb crystal NaCu$_3$Te$_2$ hosts triply degenerate nodal points (TDNPs) which are perfectly separated from the bulk states.
We find a tunable phase transition between TDNPs and a weak TI triggered by a symmetry-allowed perturbation, and we further reveal the crucial role played by the spin-orbital coupling (SOC) for the emergence of the TDNPs in this system.
Such topological non-trivial ternary compound not only serves as a perfect candidate for studying three-component fermions, but also provides a beautiful playground for understanding the topological phase transitions between TDNPs, TIs and trivial insulators, which distinguishes this system from other TDNP candidates. 
\end{abstract}

\maketitle

\section{introduction}
During the last decade, the intensive study of topological materials~\cite{RevModPhys.82.3045, RevModPhys.83.1057}, in particular the topological semimetals (SMs)~\cite{PhysRevB.85.195320, PhysRevB.88.125427, 0953-8984-28-30-303001,zhongkaicd,Xu294,PhysRevB.83.205101,PhysRevX.5.011029,Huang2015,Xu613,PhysRevX.5.031013,weyl-natphy}, has significantly advanced our understandings of the band crossings and band connectivities in condensed-matter physics. 
Around the band degeneracy points, the low-energy excitations known as quasiparticles mimic the elementary particles in quantum field theory. 
Among them, the two-dimensional (2D) Dirac fermions on the surface of TIs~\cite{Kane2005, Qi2011RMP,hgte1, hgte2,Haijunbise,bise2,bite,Chen:2010bs}, the three-dimensional (3D) Dirac fermions in the bulk of topological Dirac SMs (TDSMs)~\cite{PhysRevB.85.195320, PhysRevB.88.125427,zhongkaicd,Xu294}, the Weyl fermions in topological Weyl SMs (TWSMs)~\cite{PhysRevB.83.205101,PhysRevX.5.011029,Huang2015,Xu613,PhysRevX.5.031013,weyl-natphy},  and the Majorana fermions in the topological superconductors~\cite{Mourik1003,Nadj-Perge602,He294} directly match to their counterparts in high-energy physics. 

The quasiparticle excitations in the free fermion electronic structures can be characterized by the dimensionality of the band crossings, which allows the emergence of exotic fermions that do not exist in standard model. 
For this reason, there has been growing interest in the search for new topological SMs protected by the crystalline symmetry beyond what are ranked as 2- and 4-component fermions~\cite{PhysRevLett.115.036806,hourglass,nodalchain,PhysRevLett.116.186402}. 
It was found that 3-, 6- and 8-fold band degeneracies can emerge from non-symmorphic space groups at high-symmetric points~\cite{Bradlynaaf5037,hourglass,nodalchain,PhysRevLett.116.186402,Parameswaran}. 
Moreover, the 3-component fermions ({\it i.e.} triply degenerate nodal points (TDNPs)) can arise from symmorphic space groups as well~\cite{PhysRevX.6.031003, Weng_Topological_2016,Chang_Nexus_2017,yanghaotriple}, 
which were experimentally confirmed in MoP~\cite{Lv_Observation_2017,shekhartriple} and WC~\cite{2017arXiv170602664M}. 

\begin{figure}
\centering
\includegraphics[width=\linewidth]{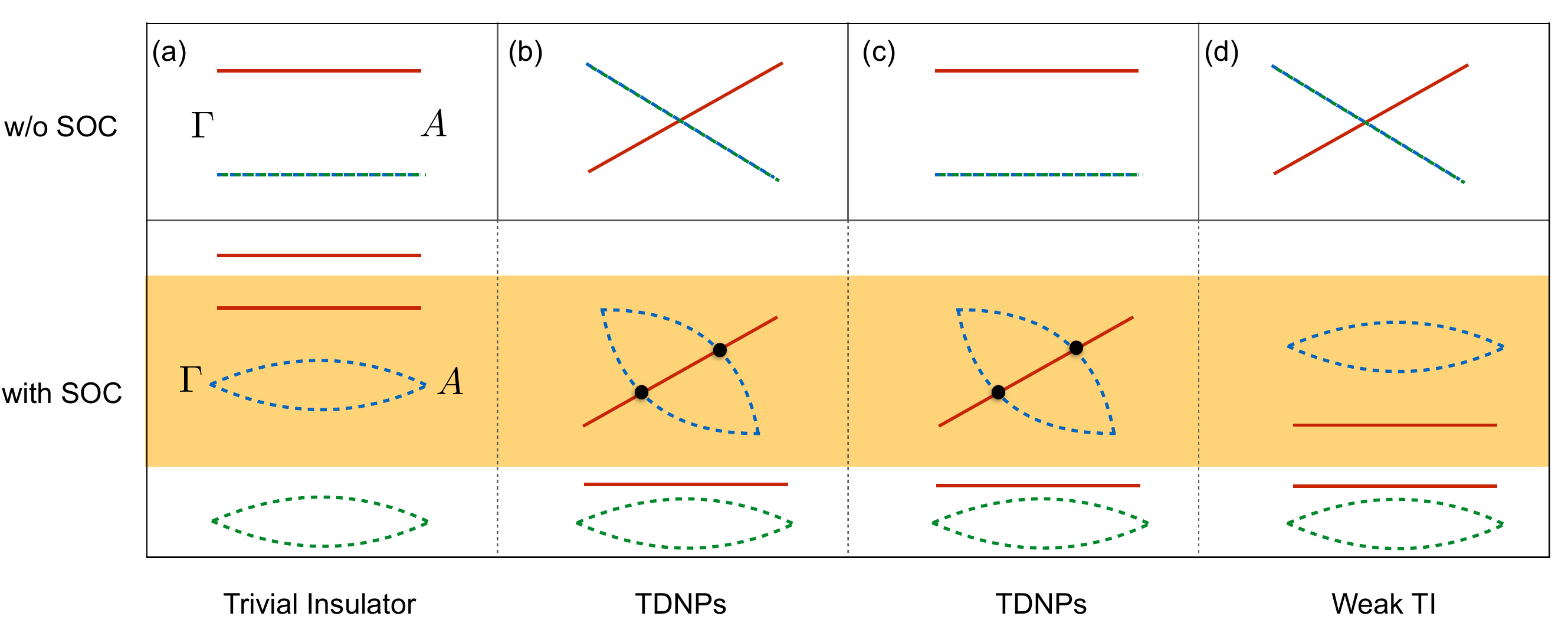}
\caption{(Color online) \textbf{Schematic plot for the TDNPs and the weak TI phases emerging from $C_{3v}$ point group symmetry. }
$\Gamma$-A supports both 2D representation (red solid line) and 1D representations (blue and green dashed lines) with the former staying above being the normal order. 
\textbf{(a)} Without inversion between the doubly and singly degenerate bands, SOC may lead to a trivial insulator.
\textbf{(b)} With the band inversion at $\Gamma$ present already at the case of w/o SOC, two singly degenerate bands cross with the doubly degenerate band leading to the appearance of a pair of TDNPs.
\textbf{(c)} SOC induces an inversion between the doubly and singly degenerate bands at $\Gamma$, resulting in the same TDNPs phase as in \textbf{(b)}.
\textbf{(d)} If the doubly and singly degenerate bands are already inverted at $\Gamma$ without SOC, the SOC can also lead to another inversion at $A$ and a full band gap between the inverted bands, which results in a weak TI.}
\label{Fig1}
\end{figure}

For TDNPs stably residing at momentum $\mathbf{k}$ in the Brillouin zone (BZ) of a symmorphic group, it is necessary for the little group to contain both one- and two-dimensional irreducible representations (IRs), which is satisfied by a 3-fold rotational symmetry $C_{3}$ and a mirror symmetry in point group $C_{3v}$~\cite{PhysRevX.6.031003} as shown in Fig.~\ref{Fig1}.   
If there is no crossing, the doubly degenerate bands (red solid line) stay above the two singly degenerate bands (blue and green dashed lines) as the normal band ordering. Spin-orbital coupling (SOC) splits each band and results in a trivial band insulator as the case of Fig.~\ref{Fig1}(a).
If the doubly degenerate bands and the singly degenerate bands cross (see Fig.~\ref{Fig1}(b)), their crossing will be stable against SOC as they belong to different IRs. 
A pair of TDNPs will appear along $\Gamma$-A, which can be viewed as the intermediate phase between TDSMs (achieved by imposing inversion symmetry) and TWSMs (achieved by breaking $C_{3v}$ symmetry)~\cite{PhysRevX.6.031003}. 
The so far discovered material candidates~\cite{PhysRevX.6.031003, Weng_Topological_2016,Chang_Nexus_2017, Lv_Observation_2017, 2017arXiv170602664M} for TDNPs belong to this class. 

However, as learned from the study of TIs, what the SOC does is more than splitting bands.
It can change a trivial band insulator into a TI by inducing band inversions.   
Therefore it is natural to expect the TDNPs to emerge in a system with normal band order solely from the relativistic effect.
As shown in Fig.~\ref{Fig1}(c), the doubly and singly degenerate bands stay apart when there is no SOC, and become crossed with a band inversion generated at $\Gamma$ when the SOC is introduced. 
In this case, the TDNPs emerge from an semimetal-insulator transition (SMIT). 
Note that the resulting TDNPs phase has no difference to the one shown in Fig.~\ref{Fig1}(b), but the driving mechanisms are distinct.
Moreover, if the band inversion strength, which is tunable by electron hopping and band dispersions, at $\Gamma$ is strong enough and the two bands at $A$ are close to each other, the SOC may also result in a new band inversion at $A$, featuring a weak TI phase shown in Fig.~\ref{Fig1}(d). 
The joint role played by the band dispersing and the SOC serves as a tunable parameter to trigger the transitions between the phases outlined in Fig.~\ref{Fig1}.
By far, the topological phase transition within the identical TDNPs system has been scarcely mentioned, let along materialized in real systems. Likewise, whether the SMIT could be achieved in real TDNPs materials remains an interrogative sentence. 

In this paper, we show the above-outlined intriguing topological phase transitions and SMIT can be realized in one single material system. 
We demonstrate that the transition between weak TIs and TDNPs could be driven by symmetry-allowed perturbations, which are experimentally highly feasible. 
The SOC driven phase transition from topologically-trivial insulator to TDNPs is realized. 
The TDNPs reside at the Fermi level, thus the system is an ideal topological triple point semimetal (TPSM), in contrary to the other discovered TDNPs, most of which sit in metals~\cite{PhysRevX.6.031003, Weng_Topological_2016,Chang_Nexus_2017,Lv_Observation_2017,2017arXiv170602664M}.
The formation mechanism of TDNPs here distinguishes from most other reported TSMs~\cite{PhysRevB.85.195320, PhysRevB.88.125427,zhongkaicd,Xu294,PhysRevB.83.205101,PhysRevX.5.011029,Huang2015,Xu613,PhysRevX.5.031013,weyl-natphy,PhysRevX.6.031003, Weng_Topological_2016,Chang_Nexus_2017,Lv_Observation_2017,2017arXiv170602664M}: it is resulted jointly from the SOC and the local couplings. 
 Additionally, in all the cases holding these different kinds of phases, the $k_{z}=0$ plane can be always viewed as a QSH insulator.

\section{Results}

\begin{figure}[htbp]
\centering
\includegraphics[width=\linewidth]{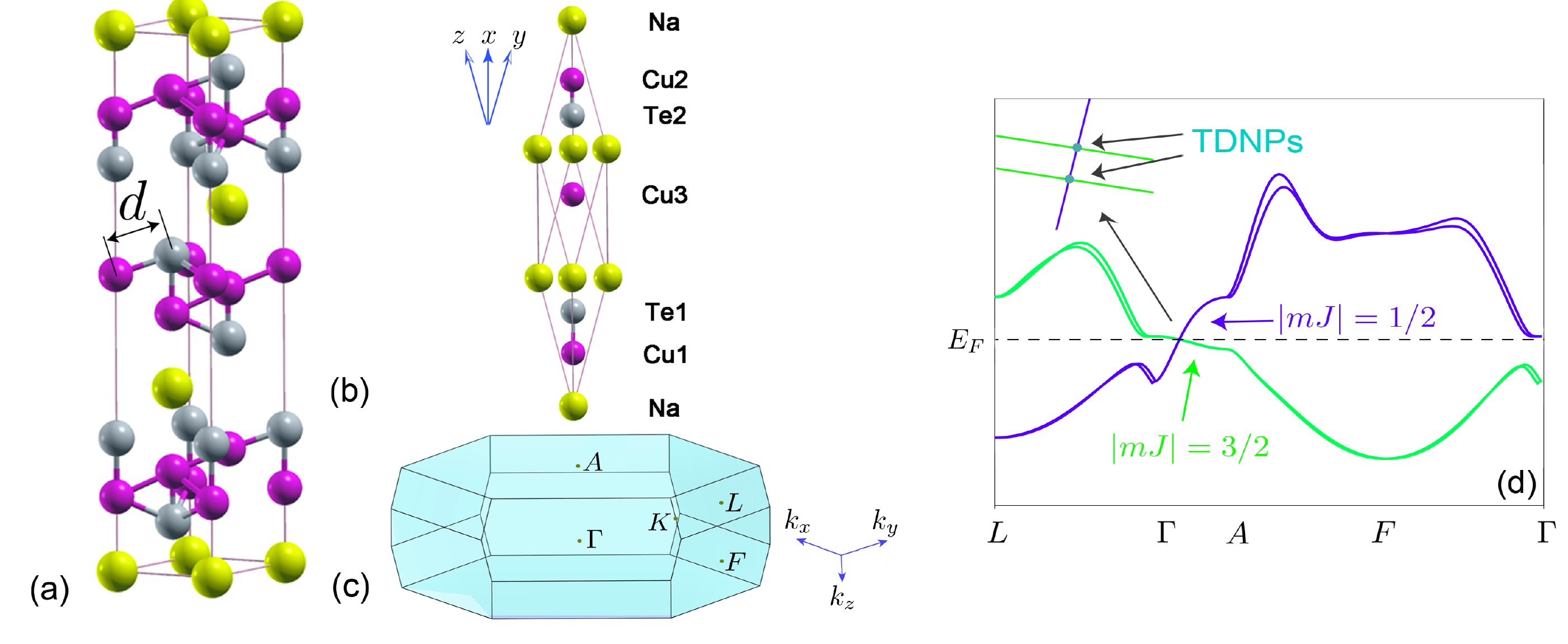}
\caption{(Color online) \textbf{Crystal and electronic structure of honeycomb NaCu$_{3}$Te$_{2}$.} \textbf{(a)} Conventional cell. \textbf{(b)} Primitive cell. \textbf{(c)} Brillouin zone. \textbf{(d)} Between $\Gamma$ and $A$, the two top valance bands and the two bottom conduction bands of the fully structure-optimized NaCu$_{3}$Te$_{2}$ admit $|mJ| = 1/2$ (blue lines) and $|mJ| = 3/2$ green lines). Their crossings are TDNPs, which are more clearly seen in the zoom in plot of \textbf{(d)}. }
\label{Fig2}
\end{figure}

\textbf{Material candidate of TDNPs.} The exotic behaviors discussed above are materialized in the hexagonal crystal NaCu$_3$Te$_2$ (Fig.~\ref{Fig2})~\cite{NaCuTe_structure}, belonging to the space group $R3m$ (160). Containing three-fold rotation $C_{3v}$ and three  mirror operations $\sigma_{\nu}$.
Along the principle axis of $C_{3}$, the atoms Na, Cu1, Te1, Cu3, Te2, Cu2 are strung in line sequently in its rhombohedral primitive cell (Fig.~\ref{Fig2}(b)). The relative position of the neighboring Cu and Te atoms (denoted by $d$) is what we have mentioned to be crucial to the electronic structure of the system, by which the band dispersion along $\Gamma-A$ is highly tunable. 
Note that Te1 and Te2, Cu1 and Cu2 are close to inversion symmetric, while the off center Cu3 eliminates inversion symmetry from the system. 

Fig.~\ref{Fig2}(d) displays the top two valence and the bottom two conduction bands of NaCu$_3$Te$_2$ calculated in a structure model fully relaxed from an initial experimental structure~\cite{NaCuTe_structure}. Other bands from higher binding energies have been dropped off for simplicity. 
Along $\Gamma-A$, these four bands resemble the situation shown inside the yellow region of Fig.~\ref{Fig1}(b). 
The band connecting the valence at $\Gamma$ and the conduction at $A$ mainly consists of Cu3-$s$, Te2-$s$ and Cu2-$d_{z^{2}}$ orbitals.
Under the SOC, $d_{z^{2}}$ orbital contains two $|mJ| = 1/2$ states which do not split under $C_{3v}$. Thus, this band remains doubly degenerate when the SOC is included. 
While, the band connecting the conduction at $\Gamma$ and the valence at $A$ consists of Cu1-$d_{xz}+d_{yz}$, Cu2-$d_{x^{2}-y^{2}}+d_{xy}$ and Te1-$p_{xy}$ orbitals. 
Their total angular momentum is $|mJ|=3/2$, making the two states non-degenerate under the SOC. 
As states with different $|mJ|$ do not hybridize, the crossing of these two bands then give rise to the pair of TDNPs shown in Fig.~\ref{Fig2}(d).  
As clearly shown, the TDNPs in NaCu$_3$Te$_2$ stay at the Fermi level which are not interfered in energy with other bulk bands. 
Thus NaCu$_{2}$Te$_{2}$ may provide by far the best material platform to study the transport anomalies and spectroscopic responses resulted from the 3-component fermions. 

\begin{figure}[htbp]
\centering
\includegraphics[width=\linewidth]{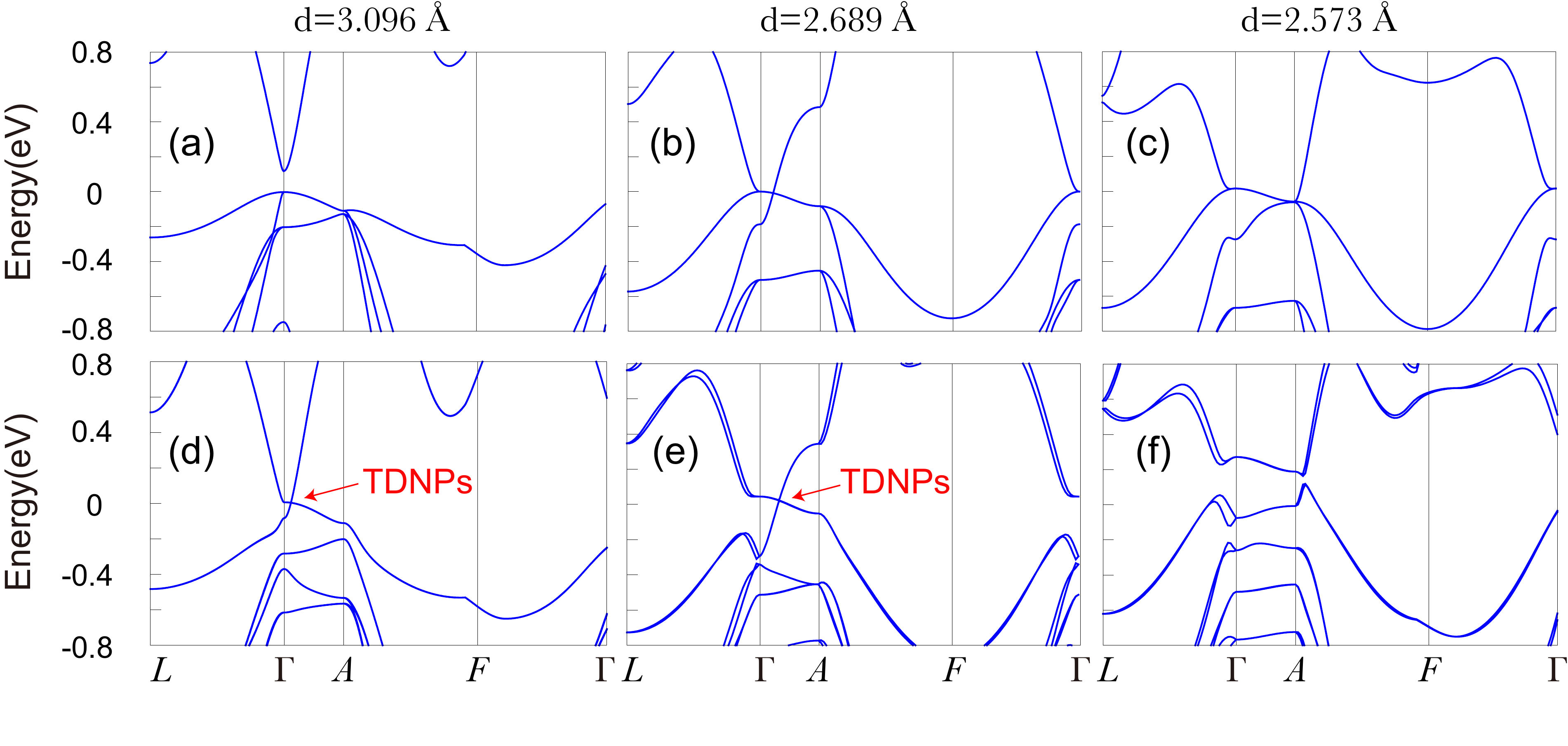}
\caption{(Color online) \textbf{The electronic structure evolution of NaCu$_{3}$Te$_{2}$ under the change of distance $d$ between Cu and Te atoms.}
The band dispersions shown in \textbf{(a/d)}, \textbf{(b/e)} and \textbf{(c/f)} correspond to the structures with $d=3.096\AA$, $2.689\AA$ and $2.573\AA$ without/with SOC, respectively. Note that the bands shown in \textbf{(c/f)} are obtained from the experimental structure~\cite{NaCuTe_structure}.}
\label{Fig3}
\end{figure}

\textbf{Topological phase transitions.} In addition to theoretically proposing a perfect material candidate for the TDNPs, we further show that this system is close to the boundary of two topologically distinct phases shown in Fig.~\ref{Fig1}(c) and (d). 
And remarkably, symmetry-allowed perturbations easily drive one phase to another, and vice versa. 
The high tunability of the topological phases in NaCu$_{2}$Te$_{2}$ is a hallmark of this system distinguishing from other TDNPs material candidates. 
In Fig.~\ref{Fig3} we show the evolution of the band structure as a function of the distance $d$ between the neighboring Cu and Te atoms without and with SOC. 
Changing the relative distance of Cu/Te along the $C_{3}$ rotational axis does not violate $C_{3v}$ symmetry, thus it is a symmetry-allowed perturbation to the system. The structures with $d=3.096 \AA$, $2.689 \AA$ corresponding to Fig.~\ref{Fig3}(a,b,d,e) are obtained from theoretical optimization. Fig.~\ref{Fig3}(c,f) ($d=2.573 \AA$) correspond to the pristine experimental  structure~\cite{NaCuTe_structure}.
The coupling of Cu with Te, effectively characterized by the value of $d$, leaves a strong influence on the band dispersion along $\Gamma-A$. 
When Cu and Te stay far apart, the singly and doubly degenerate bands separate as well and they stay in their normal order, see Fig.~\ref{Fig3}(a). 
The increased coupling between Cu and Te disperses the band between $\Gamma$ and $A$, eventually moving the doubly degenerate bands reversely below the singly degenerate states at $\Gamma$, causing an inverted band order here (Fig.~\ref{Fig3}(b)).
For the same reason explained before, the SOC will result in a pair of TDNPs in this case (Fig.~\ref{Fig3}(e)). 
Increased coupling shown in Fig.~\ref{Fig3}(c) further enhances the band inversion at $\Gamma$. 
Simultaneously, the two bands at $A$ get closer to each other and eventually stick together. 
With the SOC taken into account in this case, a new band inversion at $A$ is easily induced yielding an even number of band inversions in the system, as a result NaCu$_{3}$Te$_{2}$ is driven into a weak TI phase (Fig.~\ref{Fig3}(f)). 
We note that the critical value $d=2.657 \AA$ for the TDNPs and weak TI phase transition (see the Supplementary Information) is very close to the experimental value $d=2.573 \AA$.
This intriguing fact indicates that the experimental-realized system is close to the critical point for the topological phase transition. The material could therefore be expected to be on either side of the phase diagram under different synthesis conditions.
Additionally, the trend of the band structure evolution presented in the Supplementary Information gives the idea that the bulk gap of TI is highly tunable for the same reason. 

In addition to the coupling strength between Cu and Te, the SOC jointly plays a pivotal role for the emergence of TDNPs in this system.
In nearly all the TDNPs systems reported so far, the bulk bands neglecting SOC are already a SM. After bringing SOC into play, the crossing bands split and each crossing point is tore apart into two TDNPs, with the inversion of crossing bands might being reinforced as an additional effect. Contrarily, by comparing Fig.~\ref{Fig3}(a,d) and (b,e) one would notice that, in NaCu$_3$Te$_2$, the TDNPs can emerge from a topologically-trivial insulator solely by SOC. 
We further confirm this by calculating the electronic structure with varied strengths of the SOC $\lambda$, see Supplementary Information for more details.

\section{Discussion}

\textbf{Topological nature and surface states (SS).} The TDNP and TI phases discovered in NaCu$_3$Te$_2$ are topologically nontrivial as proven by the Wilson loop methods~\cite{Soluyanov2011,Yu2011}, see Supplementary Information. 
In both phases, the doubly and singly degenerate bands at $\Gamma$ are inverted. As the $k_{z}=0$ plane is gapped everywhere, a $Z_{2}$ topological invariant~\cite{fu1,fu2} for this plane can be defined, and it is proved to be 1. Therefore both the $k_{z}=0$ plane in the TDNP and the TI phases can be viewed as a QSH insulator characterizing the topological nature of this system.
In the weak TI phase shown in Fig.~\ref{Fig3}(f), owing to the additional band inversion at $A$ we find $Z_2=(0; 111)$,  as expected. 

\begin{figure*}
\centering
\includegraphics[width=\linewidth]{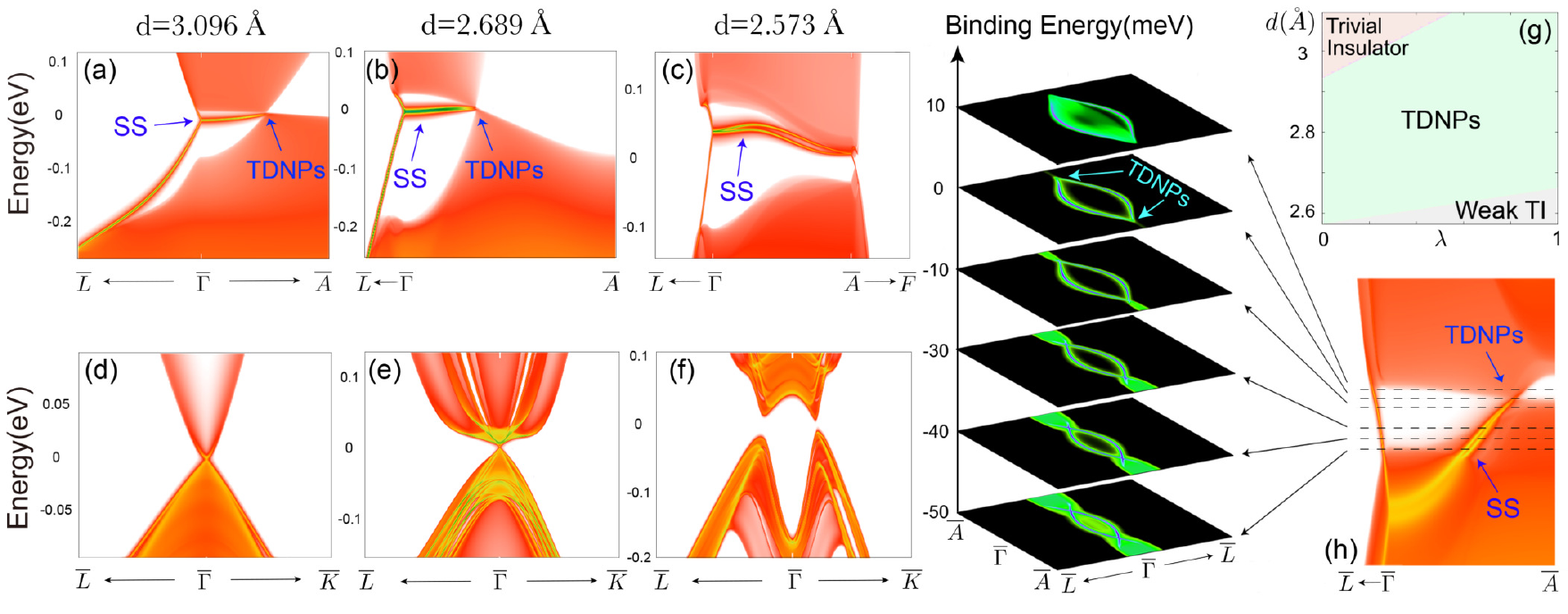}
\caption{(Color online) \textbf{TSS for the three phases discussed in Fig.~\ref{Fig3}}. \textbf{(a-c)} and \textbf{(d-f)} at the (100) and (001) surfaces of the primitive cell, respectively. \textbf{(g)} The topological phase diagram of NaCu$_3$Te$_2$ as a function of Cu-Te distance $d$ and the SOC strength $\lambda$. \textbf{(h)} The TSS on the (110) surface for $d=2.689\AA$. The six iso-energy cuts show the evolving behavior of TSS connecting the TDNPs and the surface Dirac cone at $\Gamma$. }
\label{Fig4}
\end{figure*}

Because of the non-trivial band order, the topologically protected metallic surface states are reasonably expected. Combining the tight-binding (TB) Hamiltonian~\cite{Marzari1997} and iterative method based on surface Green's function~\cite{Sancho1985}, we obtain the topological surface states (TSS) for the (100) and (111) surfaces. 
In the same order as Fig.~\ref{Fig3}, the left three columns of Fig.~\ref{Fig4} correspond to the TDNP phases formed from a trivial band insulator (Fig.~\ref{Fig4}(a,d)) and a SM (Fig.~\ref{Fig4}(b,e)), the weak TI phase (Fig.~\ref{Fig4}(c,f)), respectively.  
At (100) surface of the two TDNP phases, owing to the band inversion at $\Gamma$ TSS appear at surface $\bar{\Gamma}$ and they are connected to the SS coming from the TDNPs. 
At (111) surface, the bulk $\Gamma$ point and the TDNPs project to the same surface $\bar{\Gamma}$ so that no SS are observed. 
Similarly, in the weak TI phase, at (100) surface the two surface Dirac cones at $\bar{\Gamma}$ and $\bar{A}$ can be clearly seen, with the even number of surface Dirac cones consistent with the weak TI nature. 
At (111) surface, there is no SS as the two band inversions at $\Gamma$ and $A$ project to the same surface $\bar{\Gamma}$ point.

Based on our analysis, a general phase diagram of NaCu$_{3}$Te$_{2}$ is predicted in Fig.~\ref{Fig4}(g). We propose the effective coupling between Cu and Te atoms (which can be changed by the distance $d$ of them) and the SOC strength $\lambda$ to be the effective tuning parameters, which can efficiently drive the topological phases from one to another. 
Based on the different experimental conditions, we believe both the TDNPs and the weak TI phase can be obtained.  

In Fig.~\ref{Fig4}(h) we further show the local density of states on the (110) surface of the primitive cell, corresponding to the case of $d=2.689\AA$ (Fig.~\ref{Fig4}(b,e)), with six different iso-energy cuts to illustrate the evolution of the TSS. 
As a characteristic behavior of this system, the TSS resulted from the QSH states at $k_{z}=0$ plane are always connected to the SS originated from the TDNPs. 
The connecting point of these two SS are different on different surfaces. 
On the (100) surface, the SS from the TDNPs are nearly flat in energy and the connecting point of the two SS is well positioned inside the bulk energy gap at $\Gamma$.
In contrast, at (110) surface the connecting point is energetically embedded in the valence spectrum. 
Thus, as the decrease of the binding energy the enclosed area of the TSS shrinks.    

In summary, we proposed that topologically distinct phases, {\it i.e.} the TDNPs and the weak TIs, and their phase transitions can be realized in one single system. 
The transition between them can be easily triggered by experimentally feasible and symmetry allowed perturbations. 
We made material-specific predictions for the TDNPs in NaCu$_3$Te$_2$, which reside exactly at the Fermi level.  
We identified two phase transitions: one between the weak TIs and the TDNPs induced by local couplings, and one between topologically-trivial insulators and the TDNPs induced purely by the SOC.
The TDNPs with the resultant SS in NaCu3Te2 are awaiting to be observed by angle-resolved photoemission spectroscopy (ARPES).
We believe this stoichiometric compound can serve as a ideal material platform to study the TDNP semimetal and the related topological phase transitions, and facilitate promising applications.

\section{Methods}
   \textbf{Methods} The NaCu$_{3}$Te$_{2}$ was calculated with density functional theory. The projector-augmented-wave method implemented in the Vienna Ab-Initio Simulation Package{\cite{kresse1996}} was employed, with the exchange-correlation functional considered in the generalized gradient approximation potential~\cite{perdew1996}. By (not) treating the semi core $p$ states of Na and Cu as valence bands and applying an energy cutoff of 520 (500) eV in 84 (512) $\mathbf{k}$-points per reciprocal atom we obtain the optimized structure with $d=2.689 \AA$ ($d=3.096 \AA$).    The TSS were calculated by applying the iterative Green's function approach~\cite{Sancho1985} based on the maximally localized Wannier functions~\cite{Marzari1997} obtained through the VASP2WANNIER90~\cite{Mostofi2008}  interfaces in a non-self-consistent calculation.

\section{Acknowledgements}
The authors thank Y.-L. Chen, Z.K. Liu, and Y.F. Guo for helpful discussions.  G. Li acknowledges the starting grant of ShanghaiTech University.




\bibliographystyle{apsrev4-1}
\bibliography{ref}  

\end{document}